\documentclass[12pt]{amsart}
\usepackage{amsmath}
\usepackage{amsbsy}
\usepackage{graphicx,amsmath,amssymb,fullpage,amsfonts,bbm,bbold}
\usepackage{bbm} 

\usepackage[nodisplayskipstretch]{setspace}
\usepackage{lscape}

\usepackage{longtable}
\DeclareMathOperator{\iid}{\stackrel{\mbox{\tiny iid} }{\sim}}

\newcommand{\B}{\mathbf{B}}
\newcommand{\D}{\mathbf{D}}
\newcommand{\A}{\mathbf{A}}

\newcommand{\bS}{\boldsymbol{\Sigma}}
\newcommand{\btheta}{\boldsymbol{\theta}}
\newcommand{\x}{\mathbf{x}}
\newcommand{\Z}{\mathbf{Z}}

\newcommand{\y}{\mathbf{y}}
\newcommand{\E}{\mbox{E}}

\newcommand{\z}{\mathrm{z}}
\newcommand{\f}{\mathrm{f}}
\newcommand{\N}{\mbox{N}}
\newcommand{\bdelta}{\boldsymbol{\delta}}
\newcommand{\I}{\mathbf{I}}
\newcommand{\Psitwo}{\boldsymbol{\Psi}^2}
\usepackage{natbib}

\usepackage{lineno}

\title{Shrinkage priors for linear instrumental\\ variable models with many instruments}
\author{P. Richard Hahn$^1$}
\thanks{1. Booth School of Business, University of Chicago}
\author{Hedibert Lopes$^2$}
\thanks{2. INSPER --- Institute of Education and Research}
\thanks{The first author thanks the Booth School of Business for supporting this research.}
\date{}                                           
\begin{document}
\maketitle

\begin{abstract}
This paper addresses the weak instruments problem in linear instrumental variable models from a Bayesian perspective.   The new approach has two components.  First, a novel predictor-dependent shrinkage prior is developed for the many instruments setting.  The prior is constructed based on a factor model decomposition of the matrix of observed instruments, allowing many instruments to be incorporated into the analysis in a robust way. 

Second, the new prior is implemented via an importance sampling scheme, which utilizes posterior Monte Carlo samples from a first-stage Bayesian regression analysis.  This modular computation makes sensitivity analyses straightforward.  

Two simulation studies are provided to demonstrate the advantages of the new method.  As an empirical illustration, the new method is used to estimate a key parameter in macro-economic models: the elasticity of inter-temporal substitution. The empirical analysis produces substantive conclusions in line with previous studies, but certain inconsistencies of earlier analyses are resolved.
\end{abstract}

\section{Introduction}
This paper considers the practically important problem of how to undertake an instrumental variables analysis when the instrumental variables may be only weakly predictive of the endogenous regressor. This problem is illustrated via an applied problem from monetary policy theory: estimating the elasticity of inter-temporal substitution (EIS).  The EIS is a central parameter in the theoretically optimal consumption rule.  The weak instruments problem is addressed by including an array of instruments which---in aggregate---alleviate the weak instruments phenomenon.  In adding these many auxiliary instruments, care must be taken to avoid over-fitting, which will be achieved through a powerful shrinkage prior based on ideas from factor analysis.
Using Bayesian factor models for the purpose of inducing a regression can prove problematic \citep{hahn2013partial}:  if the dominant factor structure apparent in the instruments does not predict the endogenous regressor, estimates of the first stage regression can be strongly biased to zero, exacerbating the identification problem the instruments were intended to resolve.  
What is required instead is a prior over the first-stage regression coefficients which is biased towards any obvious factor structure, but which does not collapse sharply to zero if the evident structure in the instruments appears not to be predictive of the endogenous regressor.  It is demonstrated that a prior built on this principle can be constructed in terms of an approximate low-rank decomposition of the instruments matrix.  Finally, an importance resampling approach is developed to implement the new prior in the instrumental variables setting.
\subsection{Overview}
The balance of this paper introduces a {\em factor shrinkage prior} and explores its many relations to previous methods and its application to instrumental variable models with many instruments. Specifically, Section \ref{sec:background} lays out the background and notation of Bayesian linear IV, Gaussian linear factor models, and predictor-dependent priors for linear regression.  

Despite this rich context, the basic intuition behind the new prior is quite straightforward, and is worth delineating at the outset. Begin with a linear model for a scalar response variable $x_i$:
\begin{equation*}
x_i  = \z^t_i \bdelta + \epsilon_i; \;\;\;\; \epsilon_i \iid \N(0, \sigma^2).
\end{equation*}
A factor shrinkage prior over the vector of regression coefficients $\bdelta$ is induced via the following three steps.  
\begin{enumerate}
\item  First, suppose that $\mbox{cov}(\z) = \B\B^t + \Psitwo$ is known, for $\Psitwo$ diagonal and $\mbox{rank}(\B) \ll \mbox{dim}(\z)$.  That is, suppose that the {\em factor structure} of the predictor variables is known. 
\item Next, create an {\em over-complete dictionary} by decomposing $\Z$ into i) its projection onto the column space of $\B$ and ii) the residuals arising from this projection.  Specifically, define 
$\tilde{\Z} = 
\begin{pmatrix}
\tilde{\B}^t \Z \\
 (\I - \tilde{\B} \tilde{\B}^t) \Z
\end{pmatrix},
$
where $\tilde{\B}$ is an orthonormalization of $\B$. Note that the span of $\tilde{\Z}$ is the same as for $\Z$.
\item Redefine the likelihood in terms of $\tilde{\z}$:  $x_i  = \tilde{\z}^t_i \tilde{\bdelta} + \epsilon_i \;\;\;\; \epsilon_i \iid \N(0, \sigma^2).$  Proceed with Bayesian inference under one's preferred shrinkage prior over $\tilde{\bdelta}$.
\end{enumerate}
The intuition behind this method is simply that if the derived variables $\tilde{\B}^t\Z$ are strong predictors of $x$, the shrinkage prior on $\tilde{\bdelta}$ should allow to spot this strong signal; at the same time, if these derived variables are not by themselves adequate, the residual predictors  $(\I - \tilde{\B} \tilde{\B}^t) \Z$ have still been retained.  In the former case, one has relied on the factor structure of the predictor variables to construct an approximately sparse regression problem with $p + k$ predictors, of which $k$ are dominant.  In the latter case, one has only added $k \ll p$ predictors and so one is essentially not much worse off than if fitting the unmodified regression.

This sketch has omitted many details.  For example, in practice, the factor structure is not known exactly and so must be inferred or approximated and one must choose what prior to use once $\tilde{\Z}$ and $\tilde{\bdelta}$ have been defined. Section \ref{sec:fsprior} fills in these details and demonstrates the prior's performance via a small simulation study.   Finally, one must determine how to implement this prior within an instrumental variable analysis; Section \ref{sec:sampler} introduces an efficient importance sampler for this purpose.  

Section \ref{sec:empirics} then turns to the empirical analysis and Section \ref{sec:discuss} concludes with a brief discussion.

\section{Background}\label{sec:background}
\subsection{The Bayesian linear instrumental variables model}
This section describes a simple re-parametrization of the usual Gaussian instrumental variables (IV) model.  This representation will underpin the computational approach taken later. 

The starting point of Bayesian approaches to endogenous regressors is the structural equation model
\begin{equation}\label{structural}
\begin{split}
y_i &= \beta x_i  + \epsilon_y\\
x_i & = \z_i^t \bdelta + \epsilon_x.
\end{split}
\end{equation}
where $(\epsilon_x, \epsilon_y)$ are jointly Gaussian with mean zero and covariance 
\begin{equation*}
\mbox{cov}\begin{pmatrix} \epsilon_x \\ \epsilon_y \end{pmatrix} \equiv \mathbf{S} = \begin{pmatrix} \sigma_x^2 & \sigma_{xy} \\ \sigma_{yx} & \sigma_y^2 \end{pmatrix}.
\end{equation*}
The variable $x_i$ is referred to as the treatment variable, $y_i$ is the response variable and $\z_i$ is a vector of {\it instruments}.  The unknown parameters in this model are $\beta$, $\delta$, $\sigma_x^2$, $\sigma_y^2$ and $\sigma_{xy} = \sigma_{yx}$; the parameter of interest is $\beta$.  Because of the implied covariance between $x_i$ and $\epsilon_y$, valid estimates of $\beta$ cannot be obtained from just a regression of $y_i$ onto $x_i$.

The joint distribution of the observables can be found by substitution 
\begin{equation}
\begin{split}
x_i &= \z_i^t \bdelta + \epsilon_x,\\
y_i &= \z_i^t \bdelta \beta + \beta \epsilon_x + \epsilon_y.
\end{split}
\end{equation}
A further reparametrization yields
\begin{equation}\label{reduced}
\begin{split}
x_i &= \z_i^t \bdelta + \nu_x,\\
y_i &= \z_i^t \bdelta \beta + \nu_y,
\end{split}
\end{equation}
with 
$$\mbox{cov}\begin{pmatrix} \nu_x \\ \nu_y \end{pmatrix} = \boldsymbol{\Omega} = \A\mathbf{S} \A^t $$
where $\A = \begin{pmatrix} 1 & 0\\ \beta & 1\end{pmatrix} $.  
Equation (\ref{reduced}) is referred to as the ``reduced form" equations, in contrast to (\ref{structural}), the ``structural" equations.  These various formulations invite a host of possible prior specifications.  For a discussion of common specifications, see \cite{lopes2014bayesian}.  

The focus in this paper will be on priors for $\bdelta$ when the number of instruments $p$ is large relative to the number of available observations $n$.   Priors over the remaining parameters are determined by a factorization of the likelihood based on $\epsilon_y \mid \epsilon_x \sim \mbox{N}(\alpha \epsilon_x, \xi^2)$, where
\begin{equation}\label{comp}
\alpha = \frac{\sigma_y}{\sigma_x} \rho; \;\;\; \xi^2 = (1- \rho^2)\sigma_y^2,\\
\end{equation}
with $\rho \equiv \frac{\sigma_{xy}}{\sigma_x \sigma_y}$.  The matrix $\Omega$ can be written in terms of $\beta$, $\alpha$, $\xi^2$ and $\sigma_x^2$,
\begin{equation}
\Omega = \begin{pmatrix} \sigma^2_x & (\beta + \alpha)\sigma_x^2\\ (\beta + \alpha)\sigma_x^2 & (\beta + \alpha)^2\sigma_x^2 + \xi^2
\end{pmatrix},
\end{equation}
which in turn corresponds to the following factorization of the joint likelihood over observables $(x, y)$:
\begin{equation}\label{IV}
\begin{split}
f(x,y \mid \z) &= f(y \mid x,\z)f(x \mid \z)\\
&= \mbox{N}_{y\mid x}(x \beta + \alpha(x-\z^t\bdelta), \xi^2)\times\\
& \;\;\;\;\; \mbox{N}_x( \z^t\bdelta, \sigma_x^2).
\end{split}
\end{equation}
The appearance of $\bdelta$ in both factors on the right-hand side means that observations of $(y_i, \z_i)$ allow one to disentangle $\beta$ and $\alpha$.  It is concievable, of course, that in a given applied problem one instead has
\begin{equation}
f(x,y \mid \z) = f(y \mid x,\z)f(x \mid \z) = \mbox{N}_{y\mid x}(x \beta + \alpha(x-\z^t\bdelta), \xi^2)\mbox{N}_x( \z^t\tilde{\bdelta}, \sigma_x^2),
\end{equation}
with $\tilde{\bdelta} \neq \bdelta$.  The assumption that $\tilde{\bdelta} = \bdelta$ is referred to as the instrument exclusion restriction and in general is untestable.  See \cite{conley2012plausibly} and \cite{chan2014imperfect} for approaches which weaken this assumption, yielding only partial identification of $\beta$.  In this paper,  the exclusion restriction will be assumed.

Bayesian linear IV has been studied for many years now \citep{LindleyElSayyad1968, dreze1976bayesian,geweke1996bayesian,chamberlain1996hierarchical,chao1998posterior} and remains an active area of research \citep{kleibergen2003bayesian}. For a textbook treatment, chapter 7 of \cite{BayesMarketing} is a nice resource.  The basic approach outlined above can be modified to consider non-Gaussian error terms \citep{conley2008semi}.
In the empirical illustration considered in Section \ref{sec:empirics}, $y_i$ is the quarterly change in consumption in the United States, $x_i$ is the real interest rate and $\beta$ denotes the elasticity of inter-temporal substitution. The instrument vector $\z_i$ consists of a battery of macroeconomic indicators, twice-lagged.  This formulation of the economic problem follows from a linearization of an Euler equation; see \cite{yogo2004estimating} section II (and references therein) for details.
\subsection{Gaussian factor regression models}
Given a $p$-by-$k$ matrix $\B$ and a $k$-by-$1$ vector $\f_i$, a linear factor model for the $p$-dimensional vector $\z_i$ takes the form 
\begin{equation}
\z_i  = \B \f_i + \epsilon_i 
\end{equation}
where $\epsilon_i$ is a $p$-dimensional, independent, additive error term (referred to as {\it idiosyncratic errors}). Conditional on the factors $\f_i$, the data may be viewed as realizations of an independent and identically distributed random variable.  However, the {\em latent factor scores} $\f_i$ are not observable, rather they are given a prior distribution.  Integrating over the latent factors induces a  dependence structure among the observed data, in particular
\begin{equation}
\mbox{Cov}(\z_i) = \B \mbox{Cov}(\f_i) \B^t + \Psitwo,
\end{equation}
where $\mbox{Cov}(\epsilon_i) = \Psitwo$ is assumed diagonal.  

When the priors over the latent factors and the idiosyncratic errors are both Gaussian, $\f_i \iid \N(0, \I_k)$ and $\epsilon_i \iid \N(0, \Psitwo)$, the marginal distribution of $\z_i$ is also normally distributed,
\begin{equation}\label{margfactor}
\z_i \sim \N(0, \B \B^t + \Psitwo),
\end{equation}
and the model is called a Gaussian factor model.

Factor models have been a topic of research for more than 100 years.  A seminal reference is \cite{spearman1904} and \cite{bartholomew2011} is an excellent contemporary reference.  Bayesian factor models continue to see new developments, for example \cite{lopes2008spatial} and \cite{murray2013bayesian}.  Recent work considering the use of factor models in the many instruments context include \cite{groen2009parsimonious,ng2009selecting, hahn2011parameter} and \cite{kapetanios2010factor}.  Much of this previous work on factor models for IV analysis is non-Bayesian and the Bayesian treatments tend to focus specifically on asymptotic analysis under non-informative priors.  The present paper differs from these earlier approaches in considering predictor dependent priors and an importance sampling implementation. 
\subsubsection{Weak factors and weak instruments}
Factor models can be useful in a regression context owing to their ability to leverage ``side information".  To observe this phenomenon, consider a factor regression model specified as:
\begin{equation}\label{facreg}
x_i = \z^t_i\bdelta + \epsilon_i; \;\;\;\ \epsilon_i \sim \N(0, \sigma^2); \;\;\; \bdelta  = \btheta \B^t (\B\B^t + \Psitwo)^{-1}.
\end{equation}  
Suppose that $\z_i$ follows the distribution in (\ref{margfactor}) and that many observations are available from this distribution, whereas only a limited number of $x$ observations are available.  In this case, inference concerning $\bdelta$ still benefits, because the ``unlabeled" draws from (\ref{margfactor}) permit reliable inference concerning $\B$ and $\Psitwo$, which reduces the $p$-dimensional regression in (\ref{facreg}) to the problem of learning the $k$-dimensional vector $\theta$. (For a more general discussion of this idea, see \cite{Liang2007}.)

However, if the assumption in (\ref{facreg}) relating $\bdelta$ to $\B$ and $\Psitwo$ fails, the factor regression strategy can backfire, leading to insidious bias; in particular the true but unknown $\bdelta$ need not live in the span of $\B^t (\B\B^t + \Psitwo)^{-1}$.  Inferences made under an incorrect assumption of this form tend to exhibit a strong zero-bias when priors on $\btheta$ are centered at the origin.  A similar phenomenon  has long been recognized in the area of principal component analysis, where it is referred to as the ``least eigenvalue problem" \citep{jolliffe82, Cox}.  The illustration of this {\it weak factor} problem in the Bayesian linear factor model context is the topic of \cite{hahn2013partial}.

In the IV context, the weak factor problem relates intimately to the ``weak instrument" problem.  Note that when $\bdelta = 0$ the likelihood in (\ref{comp}) is non-unique in terms of the parameters $\beta$ and $\alpha$, with any combination having the same sum $\beta + \alpha$ giving equivalent likelihood evaluations.  The weak instruments problem refers then to cases where $\bdelta$ is small (but not zero), so that the likelihood is nearly flat for many combinations of $\alpha$ and $\beta$.  Therefore, strong zero-bias in $\bdelta$ due to the weak factor problem will directly impact inferences concerning $\beta$ by inducing a weak instrument scenario.  A natural way to avoid this difficulty is to work with a ``pure regression model", dealing only with a conditional model for $(x_i \mid \z_i)$ rather than for $(x_i, \z_i)$ jointly.  It is therefore natural to ask how evident structure in the predictor matrix might be incorporated into a prior of the regression coefficients.

In the applied context of this paper, factor structure in the instrument vector is plausible if one posits macroeconomic trends underlying joint movement of the various indicators.
\subsection{Predictor-dependent priors}
The idea of specifying a prior distribution over a set of regression coefficients in a way that depends on the observed matrix of predictor variables goes back at least to \cite{zellner}, where the so-called $g$-prior was introduced:
\begin{equation}
(\bdelta \mid \sigma^2, g) \sim \N(0, g\sigma^2 (\Z\Z^t)^{-1}).
\end{equation}
The $g$-prior continues to be a popular choice in the Bayesian variable selection literature \citep{Liangetal08,maruyamaGeorge2011}, due largely to the convenient closed form marginal likelihood it implies.  The $g$-prior can be motivated by specifying a regression problem in the de-correlated predictor space and using independent priors in that representation.  That is, supposing $\mbox{cov}(\z) \equiv \bS = \mathbf{L}\mathbf{L}^t$ is known and defining $\mathrm{w}_i \equiv \mathbf{L}^{-1}\z_i$ gives that
\begin{equation}
\mathrm{w}_i \sim \N(0,\I); \;\;\;x_i \sim \N(\mathrm{w}_i^t\boldsymbol{\eta}, \sigma^2);\;\;\; \boldsymbol{\eta} \sim \N(0,g\I),
\end{equation}
implies
\begin{equation}
x_i \sim \N(\z_i^t\bdelta, \sigma^2); \;\;\; \bdelta = \mathbf{L}^{-t}\boldsymbol{\eta}; \;\;\; \bdelta \sim \N(0,g\bS^{-1}).
\end{equation}
Zellner's $g$-prior follows from using an empirical plug-in estimate of $\bS^{-1}$.  The general idea of working in a rotated predictor representation has been fruitful in many contexts, for example in a model averaging capacity \citep{clyde1996prediction}.  \cite{west2003} introduces generalized-singular $g$-priors as a way to formally tie factor models to principal component regression, essentially by letting the prior on each $\psi_j^2$ approach a degenerate distribution at 0, so that ``the latent factors explain essentially all the variation in the predictors".  With no additive error, the observed data is assumed to arise as $\Z = \B\mathbf{F}$ and $\B$ can be computed (non-uniquely) via a generalized inverse. (In practice the eigenvalues of $\B$ will all be positive, but small values are set to zero.)

While \cite{west2003} expresses concern that ``a basic modelling issue arises from the explicit design-, and sample size-, dependence of the empirical factor model," the insight connecting factor models and $g$-priors can be applied ``in reverse" to ask:  is it possible to specify a predictor-dependent prior that allows for non-zero idiosyncratic variances?  That is, instead of using a dimension reduced design matrix based on the singular-value decomposition (SVD) of $\mathbf{Z}$,  it should be possible to use a true factor decomposition of $n^{-1}\mathbf{Z}\mathbf{Z}^t$.  Such a prior would benefit from the substantive bias that the response variable should associate more strongly with the communalities than the idiosyncratic errors, while directly avoiding the ``weak factor" problem by working with a pure regression model rather than a joint model.

The next section lays out the mechanics of producing such a decomposition and describes how to use this decomposition to construct a robust factor shrinkage prior.

\section{Factor shrinkage priors}\label{sec:fsprior}
If it were possible to extract latent factors governing the correlation structure in a vector of instruments, one might suppose that these factors would make ``strong" instruments.  However, such an approach is at risk of extracting the ``wrong" latent factors with respect to the desired regression, which could worsen the weak instruments problem.  This section builds a prior designed to nudge the regression towards apparent factor structure in the instruments matrix, without committing to the assumption that the endogenous regressor is independent of the instruments conditional on the factors.

The new factor shrinkage prior is built on two ideas, the Frisch decomposition of a matrix and a robust shrinkage prior called the horseshoe prior.  Sections \ref{sec:frisch} and \ref{sec:horseshoe} provide the details of this work.  Section \ref{sec:fsh} defines the new prior and section \ref{sec:simstudy1} conducts a small simulation study. 
\subsection{The Frisch decomposition}\label{sec:frisch}
The notion of ``shared factors" among vectors of measurements can be characterized in terms of an optimization problem motivated by the early work of Ragnar Frisch on ``confluence analysis" \citep{frisch}.  Specifically, given a covariance matrix $\boldsymbol{\Sigma}$, consider the following {\it rank minimization problem}:
\begin{equation}\label{frisch}
\begin{split}
\mbox{min}_{\D} &\;\;\;\;\mbox{rank}(\boldsymbol{\Sigma} - \D)\\
\mbox{s.t.} &\;\;\;\; \D \mbox{ diagonal},\\
&\;\;\;\; \boldsymbol{\Sigma} - \D \geq 0.
\end{split}
\end{equation}
If $\D^*$ is a solution to (\ref{frisch}), denote a matrix pair $(\Psitwo, \B)$ a {\it Frisch decomposition} of $\boldsymbol{\Sigma}$, if 
\begin{equation}
\Psitwo = \D^*; \;\;\; \B\B^t = \boldsymbol{\Sigma} - \D^*.
\end{equation}
By assuming $\boldsymbol{\Sigma}$ known, this problem is non-statistical in nature, yet it readily captures an intuition about what makes factor models appealing as descriptions of data.  Factor models are popular not merely because they decomposes covariance structure into a common component and an independent (diagonal) component, but because it is anticipated that this decomposition can be done parsimoniously.  Indeed, any $p$-by-$p$ covariance matrix has a $p-1$ dimensional factor representation (let $\Psitwo = \iota_p\I$ for $\iota_p$ the smallest eigenvalue of the SVD), whereas the Frisch decomposition demands that we have the most concise of all such descriptions.  

\begin{center}
\begin{figure}
\includegraphics[width=4in]{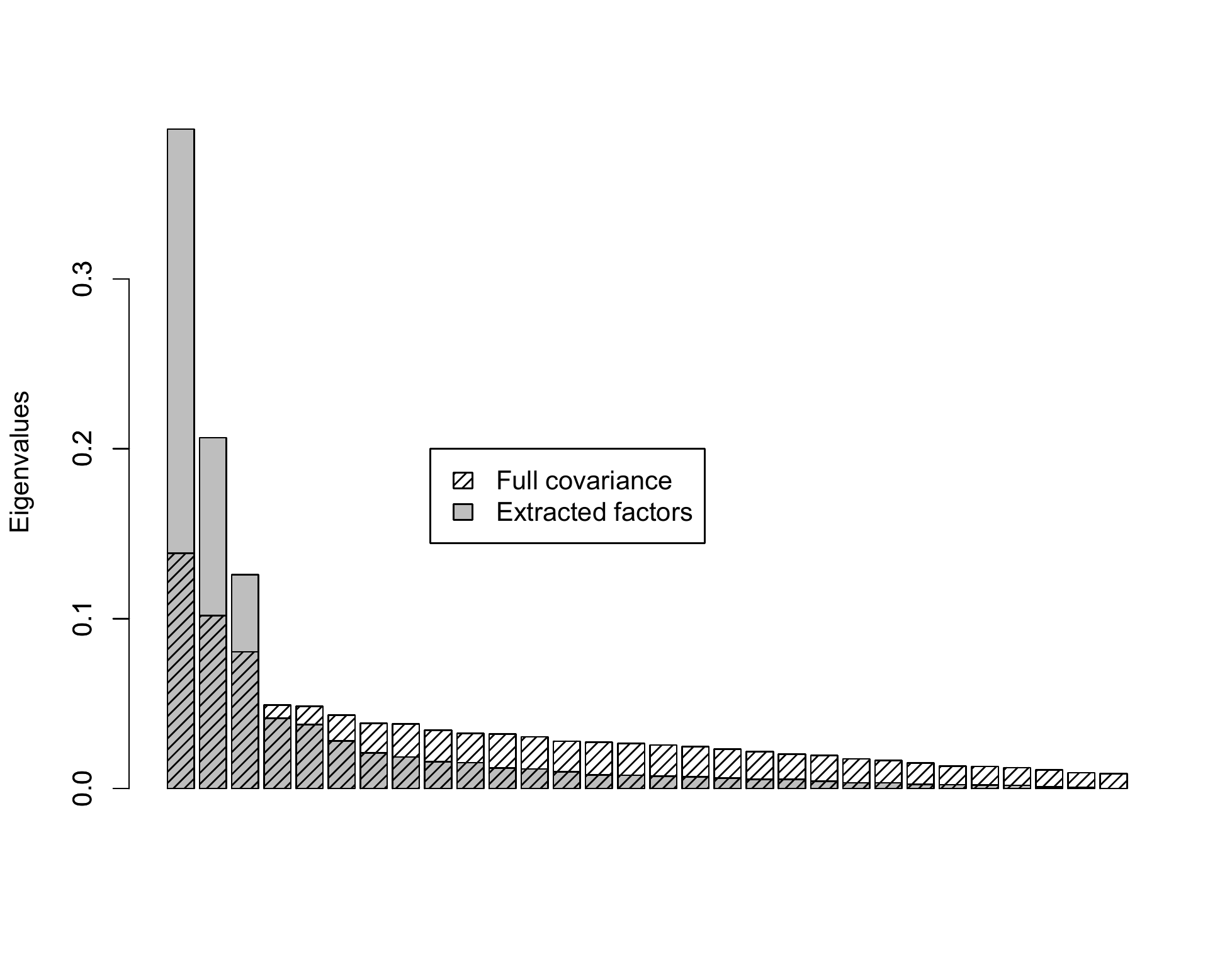}
\caption{An illustration of how the eigenvalues of a full covariance matrix $\boldsymbol{\Sigma}$  can be flatter than the eigenvalues of the trace-heuristic derived loadings matrix $\boldsymbol{\Sigma} - \D^*$ .  This occurs when $\boldsymbol{\Sigma}$ has an underlying factor structure with relatively large idiosyncratic variances.}\label{eigens}
\end{figure}
\end{center}

Alas, solving (\ref{frisch}) is quite difficult.  Fortunately, high quality approximations are available using a surrogate objective function based on the matrix  trace  \citep{fazel2002matrix}:
\begin{equation}\label{frisch_trace}
\begin{split}
\mbox{min}_{\D} &\;\;\;\;\mbox{trace}(\boldsymbol{\Sigma} - \D)\\
\mbox{s.t.} &\;\;\;\; \D \mbox{ diagonal},\\
&\;\;\;\;\boldsymbol{\Sigma} -\D \geq 0.
\end{split}
\end{equation}
The trace approximation is convex and can be routinely solved by readily available software \citep{cvx,gb08}.  The specifics of this approximation are beyond the scope of this paper; see \cite{ning2013linear} for an excellent overview with many references.  The trace approximation serves to extract a ``sharper" set of eigenvectors, in the sense of having a more rapidly decaying set of eigenvalues, as seen in Figure \ref{eigens}, which overlays the eigenvalues of an example covariance matrix $\boldsymbol{\Sigma}$  and $\boldsymbol{\Sigma} - \D^*$, where $\D^*$ solves (\ref{frisch_trace}).  In this sense, the trace heuristic still isolates ``commonalities".
Henceforth, whenever a Frisch decomposition is referred to, it is to be understood that it is computed approximately using the feasible trace formulation in (\ref{frisch_trace}).  
\subsection{The horseshoe prior}\label{sec:horseshoe}
The ``horseshoe" prior of \cite{horseshoe10} is defined as a scale mixture of normals, with representation
\begin{equation}
\pi(\delta_j) = \int{\mbox{N}(\delta_j|0,\lambda_j^2)\pi(\lambda_j^2)} d \lambda_j.
\end{equation}
To motivate this representation, consider the ``intercept only model", $(z_j \mid \delta_j) \sim \N(\delta_j, 1)$.  Then, for $(\delta_j \mid \lambda_j)  \sim \N(0, \lambda_j^2)$, the posterior mean of $\delta_j$ may be expressed as
\begin{equation}
\E(\delta_j \mid z_j) = \{1-\E(\kappa_j \mid z_j)\} z_j,
\end{equation}
where $\kappa_j = 1/(1+\lambda_j^2)$.  The prior gets its name from the fact that a half-Cauchy prior $\lambda_j \sim \mbox{C}^+(0,1)$ yields a U-shaped $\mbox{Beta}\left(\frac{1}{2},\frac{1}{2}\right)$ distribution over the ``shrinkage factor" $\kappa$, expressing the anticipation that shrinkage ought to be either severe ($\kappa \approx 1$) or minimal ($\kappa \approx 0$), and less likely to be at intermediate levels ($\kappa \approx 1/2)$.

The horseshoe and its relatives (such as \cite{Griffin12} and \cite{PolsonScott2012}) make good default priors for regression coefficients because they lack hyper-parameters and have been observed empirically to successfully shrink irrelevant coefficients strongly to zero without similarly attenuating the magnitude of relevant coefficients.
\subsection{A factor shrinkage horseshoe prior}\label{sec:fsh}
The factor shrinkage horseshoe prior arises as the implied prior on $\bdelta$ when a horseshoe prior is placed on the regression coefficients corresponding to an augmented predictor matrix.  This {\it enriched} predictor matrix is constructed using the Frisch decomposition of $\hat{\bS} = n^{-1}\Z\Z^t$ (the hat denoting that this can be thought of as a point estimate of $\mbox{cov}(\z) = \bS$).  The enriched predictor set is defined as follows.  Let $(\B, \Psitwo)$ denote the Frisch decomposition of $\hat{\bS}$ and denote by $k$ the rank of $\B$. Let $\tilde{\B}$ denote the orthonormalization of $\B$.  The enriched matrix is then defined as
\begin{equation}\label{aug}
\tilde{\Z} = 
\begin{pmatrix}
\tilde{\B}^t \Z \\
 (\I - \tilde{\B} \tilde{\B}^t) \Z
\end{pmatrix}.
\end{equation}
Note that $\tilde{\Z}$ is dimension $(p+k)$-by-$n$. Complete the regression model via 
\begin{equation}
\begin{split}
x_i & = \tilde{\z}^t_i \tilde{\bdelta} + \epsilon_i \;\;\;\; \epsilon_i \iid \N(0, \sigma^2)\\
\tilde{\bdelta} &\sim \N(0,s^2\boldsymbol{\Lambda}^2), \;\;\;\lambda_{j} \sim \mbox{C}^+(0,1), \;\;\;s \sim  \mbox{C}^+(0,1).
\end{split}
\end{equation}
The matrix $\boldsymbol{\Lambda}$ is diagonal with local shrinkage factors $\lambda_j$, $j = 1, \dots, p+k$. Denote by $\boldsymbol{\Lambda}_f$ the upper $k$-by-$k$ block of $\boldsymbol{\Lambda}$, associated with the derived factors, and $\boldsymbol{\Lambda}_r$ the lower $p$-by-$p$ block associated with the residuals.  
\subsubsection{Local shrinkage and over-complete dictionaries}
It may appear that nothing has been gained via working with the augmented design matrix. Indeed, the implied prior over $\bdelta$ under (\ref{aug}) is mostly similar to a typical regression prior.  In the special case where $\Lambda_f = \I_k$ and $\Lambda_r = \I_p$ are considered fixed, the prior on $\bdelta$ is simply a standard normal:
\begin{equation}\label{hs_fs}
\begin{split}
\Z^t \bdelta &= \Z^t \tilde{\B} \tilde{\bdelta}_f + \Z^t(\I - \tilde{\B}\tilde{\B}^t) \tilde{\bdelta}_r,\\
\bdelta & = \tilde{\B} \tilde{\bdelta}_f + (\I - \tilde{\B}\tilde{\B}^t) \tilde{\bdelta}_r,\\
\bdelta &\sim \N(0, \tilde{\B}\tilde{\B}^t + (\I - \tilde{\B}\tilde{\B}^t)(\I - \tilde{\B}\tilde{\B}^t)^t) = \N(0, \I).
\end{split}
\end{equation}
The last line follows from the idempotence of $\I - \tilde{\B}\tilde{\B}^t$.

However, the models are in fact quite different when the local hyper-variances are taken into account.  The over-parametrized augmented matrix $\tilde{\Z}$ allows an expansion of what it means to be ``local", by creating new composite predictors that are themselves linear combinations of the original predictors.  In this enriched set, the composite predictors may be found to represent the large signals, allowing more of the original predictor coefficients to be severely zero-shrunk.

The factor shrinkage prior construction suggests that local shrinkage priors combined with over-complete dictionaries could be a powerful general method for constructing novel priors for regression models.
%
%
%
\subsubsection{Computational details}
For completeness, note two additional details concerning the implemented Frisch decomposition.  First, the solution to (\ref{frisch}) is invariant to row and column scaling operations, while (\ref{frisch_trace}) is not.  This observation has motivated weighted minimum trace approximations that attempt to define and compute an optimal weight matrix \citep{Shapiro1982weighted,ning2013linear}.  As a crude heuristic, the approach taken here is to solve (\ref{frisch_trace}) applied to the sample correlation matrix as opposed to the sample covariance matrix.  

Similarly, because $\bS$ is only known up to an empirical estimate, the actual rank of $\B$ will tend not to be reduced.  Accordingly, $\tilde{\Z}$ is constructed by approximating $\B$ (respectively, $\tilde{\B}$) by its first few dominate eigenvectors.  This approximation entails that  the associated $\Psitwo$ will not be perfectly diagonal, but only ``nearly" diagonal.  

These two approximations determine the precise specification of the prior in (\ref{hs_fs}), but do not change the underlying motivation and intuition.   Moreover, the next section demonstrates that they do not demonstrably affect the qualitative behavior of the resulting posterior estimator.
\subsection{Comparison study}\label{sec:simstudy1}
This section compares the performance of the new prior to that of a full factor model and a pure regression model. Two regimes were considered, both with $p=30$ and $n=60$. In both cases data $\z_i$ is drawn from a factor model with parameters $\B$ and $\Psitwo$ generated as follows.  For $j = 1, \dots, p$ and $g = 1, \dots, k$
\begin{equation}\label{simstudy}
\begin{split}
a_{j,g} &\sim \N(0,1)\\
w_{g} &\equiv 1 + |\epsilon_g|, \mbox{ s.t. } |w_{g}| \geq |w_{g'}| \mbox{ if } g < g',\\
\epsilon_g &\sim \mbox{t}(0, df = 5),\\
\B &\equiv \mathbf{A}\mathbf{W},\\
\psi_{j}&= \sqrt{\mathbf{b}_j\mathbf{b}_j^t}/u_j, \hspace{0.1in} u_j \sim \mbox{Unif}(1/2,7/4),
\end{split}
\end{equation}
where $\mathbf{W}$ is a $k$-by-$k$ diagonal matrix with diagonal elements $w_g$.  The response variable $x_i$ is then generated from the factor model (\ref{facreg}) with $\sigma = 1/5$.  This gives a signal-to-noise ratio of 5-to-1 conditional on $\f_i$, representing a quite strong signal if the factors were observable.  From this basic procedure, two regimes are considered.  In the first regime, $k=3$, and the first and most dominant factor (in the sense of $|w_{g,g}|$ being largest) is solely predictive of $x_i$: $\btheta = (1, 0, 0)$.  In the second regime, $k=10$, and the least dominant factor is the one which is solely predictive of $x_i$:  $\btheta = (0, 0, \dots, 1).$  Simulations under each regime consisted of 500 replications.  Performance was judged using root mean square prediction error (RMSE), scaled by the theoretically best possible generalization error as determined by the simulated parameter values:
\begin{equation}
\begin{split}
\mbox{RMSE} &= \frac{\sqrt{\sigma^2 +n^{-1}\sum_i  \left (|\z_i^t (\bdelta - \hat{\bdelta})|^2 \right)}}{\sigma}.
\end{split}
\end{equation}
Under the simulation protocol described, $\sigma = \sqrt{1 - m + 1/25}$ where $m$ denotes the $(1,1)$ entry of $\mathbf{M} = \B^t (\B\B^t + \Psitwo)^{-1}\B$ under the first regime and the $(10,10)$ entry under the second.

\begin{center}
\begin{figure}
\includegraphics[width=3.5in]{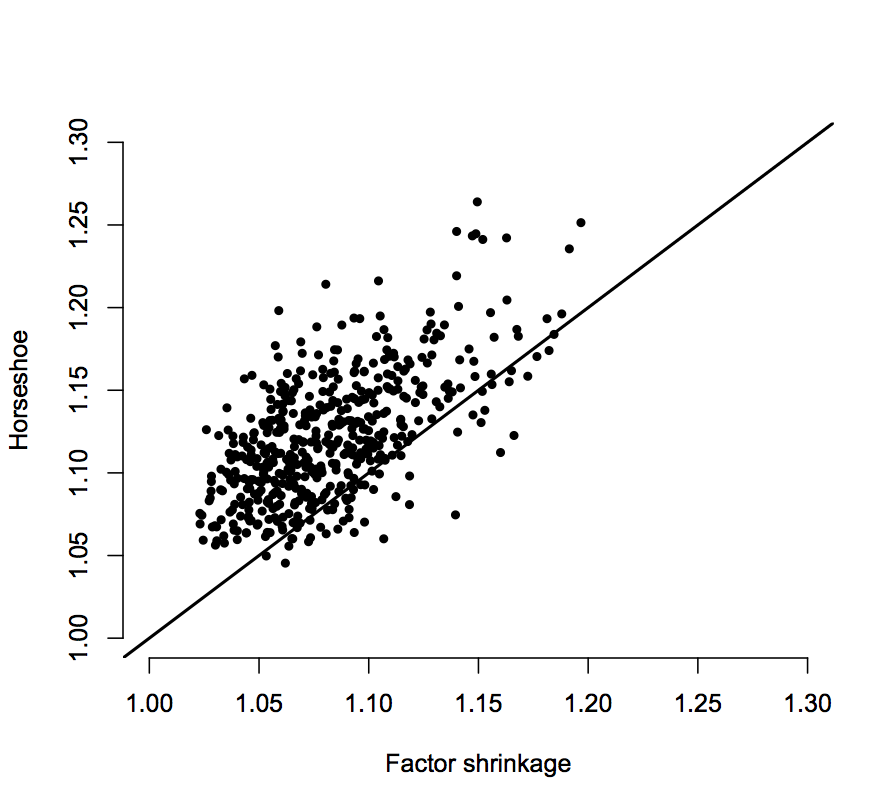}
\caption{The RMSE as a percentage of the optimal. When factor structure lay beneath idiosyncratic noise, the factor shrinkage prior dominates the unmodified horseshoe prior.}\label{mseplot}
\end{figure}
\end{center}

Intuitively, the first regime is favorable to a factor model, because $x_i$ associates strongly with the dominant factor and $n=60$ observations ought to provide information about this dominant trend of covariation. Conversely, the second regime should prove challenging for a factor model, as $x_i$ is not associated with the dominant factors; in this regime one might expect a pure regression approach to perform better.  The results in Tables \ref{simu1} and \ref{simu2} show that indeed these intuitions are borne out.  The factor shrinkage approach matches the better performing method in each case.  Figure \ref{mseplot} illustrates the benefits of the factor shrinkage in the favorable regime; not only is the average error better as reported in the tables, but it is more often the better performing method as well, indicated by the majority of the plotted points lying above the diagonal.
\begin{table}[h!]
\begin{center}\caption{Case one: when the dominant factor structure is highly predictive of the response, the factor shrinkage prior performs on par with the full factor model regression. Reported numbers are given as percent of the theoretical optimal RMSE}\label{simu1}
\begin{tabular}{lcc}
Method &  RMSE \\
 \hline
Factor shrinkage &1.09\\
Factor model  &1.09\\
Regression model &1.13\\
 \end{tabular}
\end{center}
\end{table}
\begin{table}[h!]
\begin{center}\caption{Case two: when the factor structure is less predictive of the response, the factor shrinkage approach performs on par with the pure regression model (both with horseshoe priors), while the full factor model over-shrinks. Reported numbers are given as percent of the theoretical optimal RMSE.}\label{simu2}
\begin{tabular}{lcc}
Method &  RMSE \\
 \hline
Factor shrinkage &1.17\\
Factor model  &1.28\\
Regression model &1.17\\
 \end{tabular}
\end{center}
\end{table}
\section{An importance resampler for Bayesian IV}\label{sec:sampler}
Importance sampling a Bayesian IV models proceeds analogously to two-stage least squares in that one first fits a model for $x_i \mid \z_i$ to obtain estimates of $\bdelta$.  Given $\bdelta$, estimates for $\beta$, $\alpha$ and $\xi^2$ follow straightforwardly from a regression analysis.  However, unlike two-stage least squares, which disregards the contribution of $f(x \mid y)$ in forming the first-stage estimate, a Bayesian sampling approach can account for both parts of the likelihood when obtaining posterior draws of $\bdelta$.  Integrating $\alpha, \beta$ and $\xi^2$ from the model a priori yields
\begin{equation}
\pi(\bdelta, \sigma_x^2 \mid \x, \y, \Z) \propto \pi(\bdelta, \sigma_x^2 \mid  \x, \Z) f(\y \mid \x, \Z, \bdelta),
\end{equation}
which reveals that one can obtain posterior draws from $\pi(\bdelta, \sigma_x^2 \mid \x, \y, \Z)$ by first sampling from  $\pi(\bdelta, \sigma_x^2 \mid  \x, \Z)$ as if $\y$ were not observed, and then resampling with weights proportional to $f(\y \mid \x, \Z, \bdelta)$.  Draws of $(\alpha, \beta, \xi^2)$ are then obtain compositionally, conditional on a given value of $\bdelta$.  

In the following, assume a normal-inverse-Gamma prior is used for $(\alpha, \beta, \xi^2)$, with prior mean $\E(\alpha) = \E(\beta) = 0$, covariance of $c\I$, and Gamma shape parameter of $s/2$ and scale parameter of $v/2$. Define $\tilde{x}_i \equiv (x_i, x_i - \z_i\bdelta)$. Let $\mathbf{M} = c^{-1}\I + \tilde{\x}^t \tilde{\x}$, $b = s + \y^t\y - \y^t\tilde{\x} \mathbf{M}^{-1} \tilde{\x}^t \y$, and $a = n + v$. Note that $\tilde{\x}$, $\mathbf{M}$, $a$ and $b$ depend implicitly on $\bdelta$; in particular, let subscript $j$ denote dependence on the $j$th sample of $\bdelta$.
\newpage
\begin{enumerate}
\item Draw $N$ samples of $\bdelta$ from $\pi(\bdelta, \sigma_x^2 \mid  \x, \Z)$ using the sampler described in \cite{carvalho2009handling} (though any regression model of choice will suffice here).
\item Resample with weights proportional to $f(\mathbf{y} \mid \bdelta, \x, \Z)$. Under the conjugate prior described above,
$y_i \mid z_i, x_i, \bdelta, \alpha, \beta, \sigma^2_{y \mid x} \sim \mbox{N}(x_i\beta + \alpha(x_i - \z_i^t\bdelta), \sigma^2_{y \mid x})$, for each $i$ implies that marginally over $(\alpha, \beta, \sigma^2_{y \mid x})$ the $n$-vector of responses has a multivariate $t$-distribution: $  \mathbf{y}  \mid \mathbf{x}, \Z,\bdelta \sim \mbox{t}(a, \mathbf{M})$. Therefore the resampling weights are determined for draw $\bdelta^{(j)}$ as $w_j \propto \mbox{det}(\mathbf{M}_j)^{-\frac{1}{2}}b_j^{-\frac{a_j}{2}}$.
\item Finally, sample $(\alpha, \beta, \sigma^2_x)$ given  $\bdelta$ from $\pi(\alpha, \beta, \sigma^2_x, \xi^2 \mid \mathbf{x}, \Z, \mathbf{y}, \bdelta)$, which is a conjugate Gaussian regression with predictor vector $\tilde{\mathbf{x}}$.  More specifically, draw $\sigma^2_x$ from an inverse-Gamma distribution with shape parameter $b/2$ and scale parameter $a/2$, then draw $(\alpha, \beta)$ as a vector with mean $\mathbf{M}^{-1}\tilde{\x}^t\y$ and covariance $\sigma^2_x \mathbf{M}^{-1}$.
\end{enumerate}
\subsection{Synthetic example}
This section demonstrates the efficacy of the new approach using synthetic data where the true parameters are known for post-analysis evaluation.  The intent of this exercise is not to argue that the factor shrinkage prior is better than alternatives in any absolute sense; the goal is rather to illustrate the role played by  predictor-induced bias in posterior inferences in an IV problem.

The parameters of this demonstration are set to mimic the applied analysis in the following section:  $\alpha = -0.08$ and $\beta = 0.2$.  The instruments are generated from a $k=3$ factor model as in the previous simulation.  For this demonstration, $p=20$ and $n=60$.

Two priors for the `first stage' regression coefficients $\bdelta$ are compared,  the horseshoe priors and the new factor shrinkage prior.  In the instrumental variables regression context mean squared prediction error is not the primary focus, rather it is inferences concerning the structural parameter $\beta$ that are relevant.  To reflect this inferential focus, the simulation study considers the coverage and size of the 95\% intervals produced by the two models over 250 simulated data sets.

\begin{figure}
\begin{center}
\includegraphics[width=4.5in]{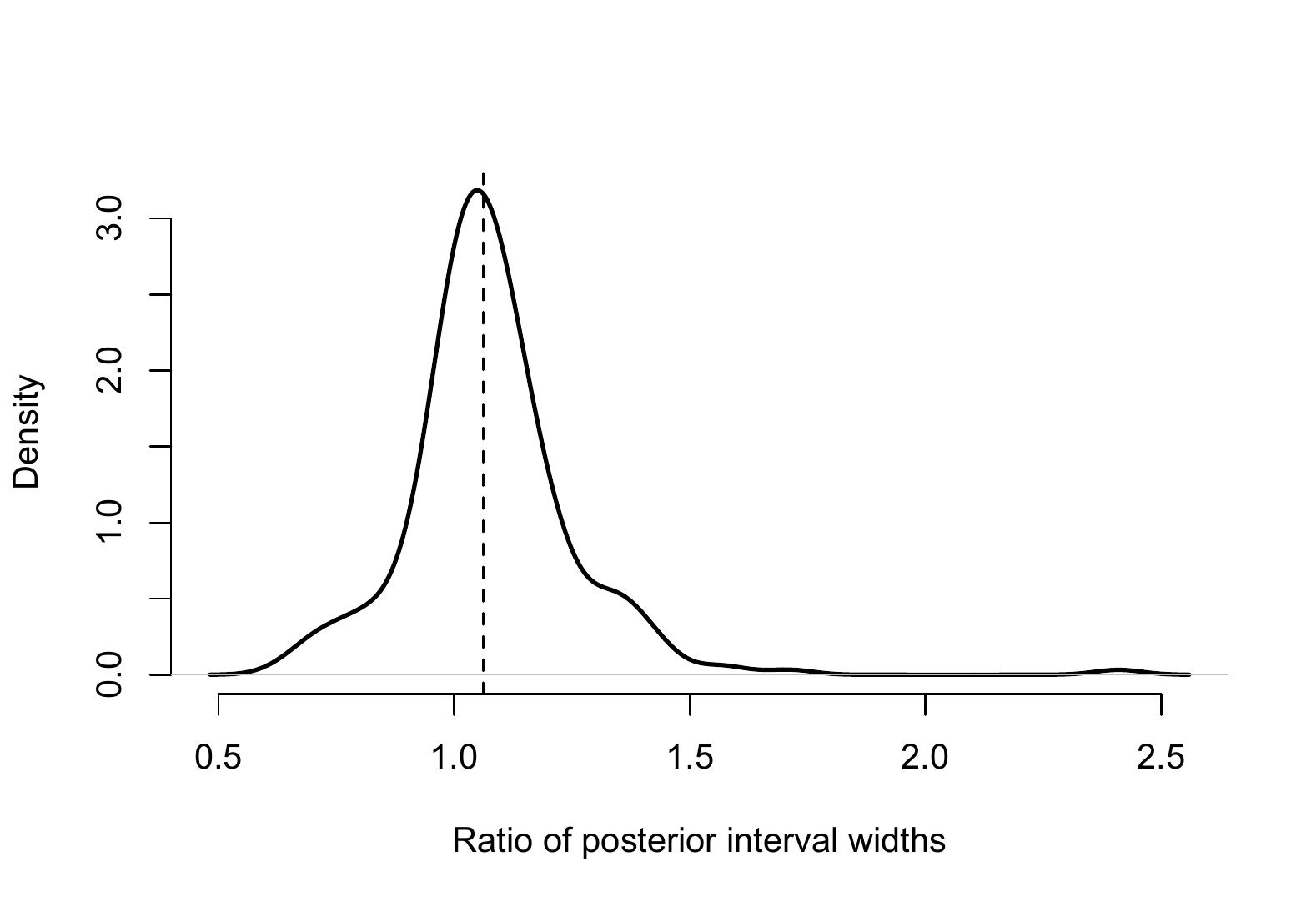}
\caption{A kernel density plot of the ratio between the posterior 95\% interval widths of the horseshoe IV regression versus the factor shrinkage IV regression.  The factor shrinkage intervals are smaller on 71\% of simulated data sets with an average decrease in interval length of 6\%.}\label{intervals}
\end{center}
\end{figure}

The upshot of the study is that the two regression methods have identical coverage of 94.8\% (237 out of 250) that is very nearly identical to the nominal coverage.  However, the factor shrinkage prior is, on average, 6\% smaller.  Figure \ref{intervals} profiles this difference via a smoothed histogram of this ratio across simulated data sets.
\section{Empirical study: the elasticity of inter-temporal substitution}\label{sec:empirics}
\cite{yogo2004estimating} considers estimating the elasticity of inter-temporal substitution via a linearization of the Euler equation,  using macroeconomic data and an instrumental variable analysis.  \cite{ng2009selecting} extend this analysis by incorporating many additional macro variables (detailed in \cite{ludvigson2007empirical}) as instruments and consolidating them into factors using a boosting approach.  This section mimics that analysis for comparative purposes, focusing on the 1970:3 to 1998:4 quarterly data for the United States.  The complete set of factors use by \cite{ng2009selecting} was unobtainable; of their 209 macro-variables a subset of 82 are used here, listed by variable code in the appendix.  A representative subset of these macro-variables includes, for example, gross domestic purchases, fixed investment in durable equipment, assets abroad, and net exports.

It is a practically relevant question as to whether or not (lagged) macroeconomic indicators serve as valid instruments in the sense of satisfying the exclusion restriction. On the one hand, under a causal interpretation it seems reasonable to assert that past indicators should only relate to the present economy via the more recent indicators---a sort of Markov property.  On the other hand, this narrative falls apart when one considers latent common causes that serve to induce dependence between today's indicators, yesterdays indicators, and today's response variable. Such shared common causes clearly violate the desired exclusion restriction. That said, this possibility will not be discussed further here; rather, a narrow comparison is drawn with the results of \cite{ng2009selecting}, who assume the validity of the macro indicators as instruments.

For reference, the model being fit is as in (\ref{IV}): $f(x,y \mid \z) =  \mbox{N}_{y\mid x}(x \beta + \alpha(x-\z^t\bdelta), \xi^2) \mbox{N}_x( \z^t\bdelta, \sigma_x^2)$,
where $y_i$ is the quarterly consumption growth (i.e., the change in consumption) in the United States, $x_i$ is the real interest rate and $\beta$ denotes the elasticity of inter-temporal substitution (EIS). The instrument vector $\z_i$ consists of aforementioned macroeconomic indicators (twice lagged), in addition to the original instruments used in\cite{yogo2004estimating}:  twice lagged nominal interest rate, inflation, consumption growth, and log dividend-price ratio.  See \cite{yogo2004estimating} section II for a theoretical justification of this model.

One goal of estimating EIS centers around  the hypothesis that it is precisely 1, which corresponds to the theoretical proposition that an investorÕs optimal consumption level is a constant proportion of wealth.  If $\beta < 1$ is less than 1, the investorÕs optimal consumption-wealth ratio is increasing in expected returns, if $\beta > 1$ it is decreasing.  

Additionally, a statistical puzzle was laid out by Yogo concerning testing the hypothesis that EIS is small.  One can estimate EIS via two distinct linearizations of the Euler equation.  Denote the estimand EIS by $\psi$.  One can estimate this directly, as described above, so that $\beta \equiv \psi$.  Alternatively, one may interchanging the response variable (consumption growth) and the regressor (real interest rate), whence $\psi \equiv 1/\beta$.  When comparing these two approaches, one often finds that both $\psi$ and $1/\psi$ are estimated to be insignificantly different than zero, which gives an apparent contradiction.

To estimate this model, a factor shrinkage prior is placed on $\bdelta$ and the conjugate normal-inverse-gamma prior described in Section \ref{sec:sampler} is used for $(\alpha, \beta, \xi^2)$, with parameters $s=1$, $v=1$, and $c = 25$.

\begin{figure}
\begin{center}
\includegraphics[width=4.5in]{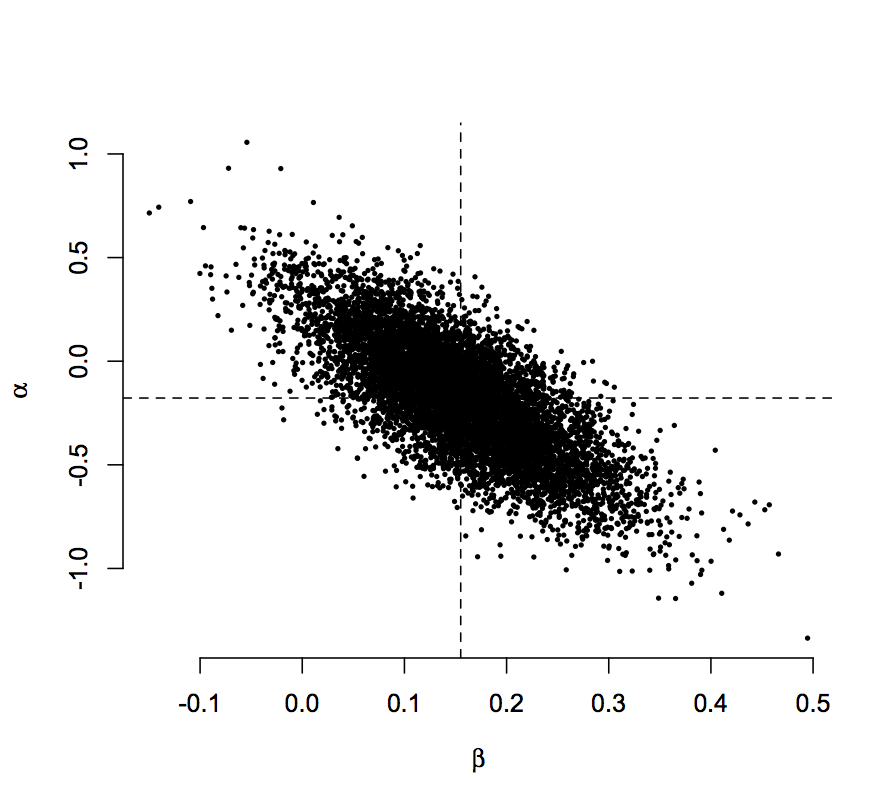}
\caption{Posterior draws of $(\alpha, \beta)$.}\label{postscatter}
\end{center}
\end{figure}

\begin{figure}
\begin{center}
\includegraphics[width=4.5in]{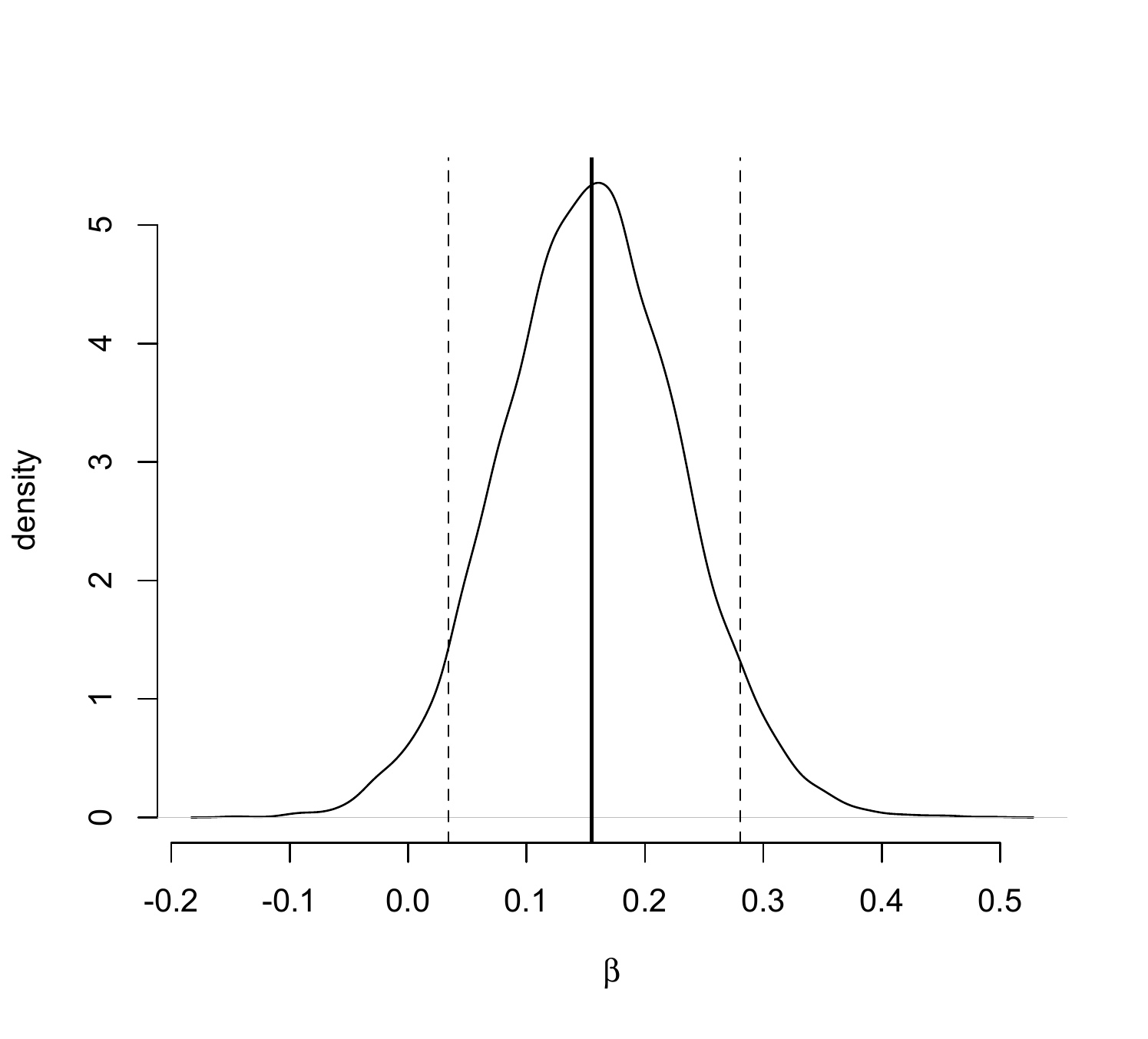}
\caption{The marginal posterior of the coefficient for elasticity of demand (here $\beta = \psi$).  The 90\% posterior credible interval does not include 1.}\label{density1}
\end{center}
\end{figure}

Using the direct form of the linearization, so that $\beta \equiv \psi$, the partial factor shrinkage IV model gives a posterior mean rate of inter-temporal substitution of approximately 16\%, with 95\% credible interval of $(1.4\%,30.8\%)$.  This is notably higher than the earlier analyses and the credible interval safely excludes 1.  Figures \ref{postscatter} and \ref{density1} summarize the posterior inference concerning $\psi \equiv \beta$. Table \ref{EIStable} compares the estimates and standard errors/posterior uncertainty for various estimation methods.  Bayesian IV with the factor shrinkage prior is the only approach which gives an estimate of $\beta \equiv \psi$ greater than the OLS estimate (0.16 versus 0.12 respectively); in particular this arises due to a posterior mean estimate of $-0.10$ for $\alpha$.

\begin{figure}
\begin{center}
\includegraphics[width=4.5in]{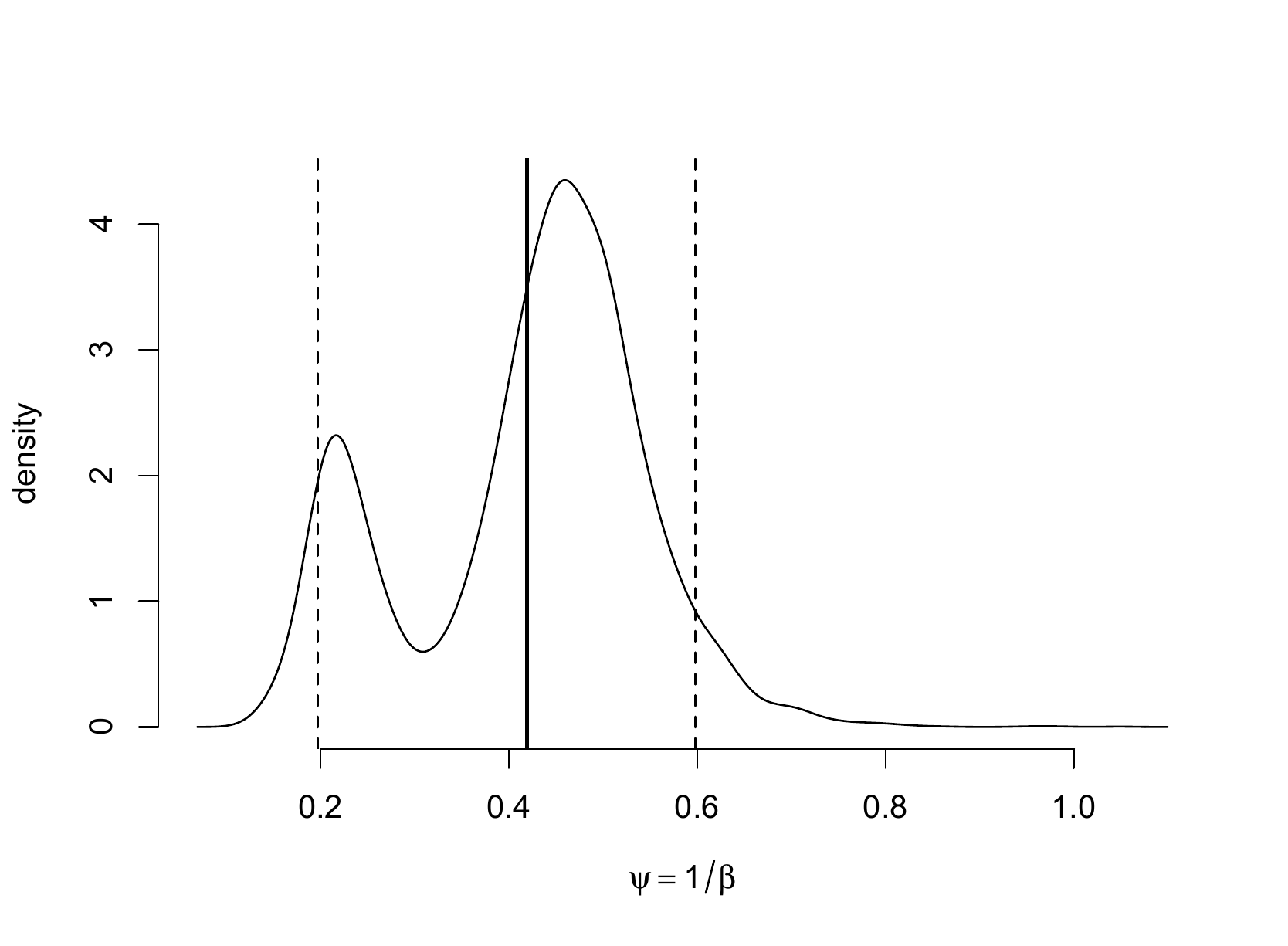}
\caption{The marginal posterior of the coefficient for elasticity of demand using the inverted regression.  The 90\% posterior credible interval for $\psi$ still does not include 1. The multimodal posterior is typical of horseshoe-based posteriors in low signal-to-noise ratio problems.}\label{density2}
\end{center}
\end{figure}

Using the inverted form of the linearization, so that $\beta \equiv 1/\psi$, the partial factor IV model gives a posterior mean for $\psi$ of 0.41, with 95\% credible interval of $(18\%, 63\%)$.  Although these estimates differ markedly from the direct regression (it {\em is} a distinct model with distinct priors), notice that no paradox emerges.  In both cases, $\psi$ is estimated to be below 1 and $1/\psi$ is estimated to be above 1. As shown in Figure \ref{density2}, however, this form of the regression has much weaker signal-to-noise ratio, which results in a multimodal posterior.  
\begin{table}[h!]
\begin{center}\caption{Estimates of the elasticity of intertemporal substitution using the direct regression ($\beta \equiv \psi$), by various methods: ordinary least squares (OLS), two-stage least squares (TSLS) for Yogo's original four instruments and for the augmented vector including the 82 macro indicators ,  Bayesian IV with factor shrinkage prior (FSP), and the boosted factor IV of \cite{ng2009selecting}, Table 7b ($\mbox{FIV}_b$). Standard errors for Bayesian models are given as the posterior standard deviation. All figures are have been rounded to two decimal places for comparison.}\label{EIStable}
\begin{tabular}{lcc}
Method &  $\hat{\psi} \equiv \hat{\beta}$& standard error \\
 \hline
OLS &0.12&0.05\\
TSLS (Yogo)  &0.06&0.09\\
TSLS (full) &0.23&0.10\\
FSP &0.16&0.08\\
$\mbox{FIV}_b$ &0.09&0.06
 \end{tabular}
\end{center}
\end{table}
\section{Discussion}\label{sec:discuss}
The factor shrinkage prior leverages an atypical matrix decomposition to create a prior that favors regression coefficients consistent with factor structure underlying the matrix of instruments.  In the language of factor analysis, this prior asserts that the treatment variable is more likely to depend on the communalities of the instrument matrix than on the idiosyncrasies.  A resampling approach is implemented which allows efficient computation and hence straightforward sensitivity analysis. Specifically, the resampling weights only require computing the determinant of a $d+1$ dimensional matrix, where $d$ is the dimension of the treatment variable (typically one), irrespective of the number of instruments. 

Analysis on synthetic data reveals that the new prior performs according to intuition:  when factor structure predictive of the treatment is apparent in the matrix of instruments, this concordance with the prior yields tighter inference concerning the treatment effect of interest ($\beta$). Meanwhile, working with a pure regression model sidesteps the least-eigenvalue problem that plagues direct factor modeling.  Moreover, the new prior can be used even when the instruments are not jointly Gaussian, such as many binary instruments. More generally, the efficacy of the factor shrinkage prior speaks to the possibilities of combining local shrinkage priors with over-complete dictionaries.

\singlespace

\bibliographystyle{abbrvnat}
\bibliography{fsp}

\begin{thebibliography}{40}
\providecommand{\natexlab}[1]{#1}
\providecommand{\url}[1]{\texttt{#1}}
\expandafter\ifx\csname urlstyle\endcsname\relax
  \providecommand{\doi}[1]{doi: #1}\else
  \providecommand{\doi}{doi: \begingroup \urlstyle{rm}\Url}\fi

\bibitem[Bartholomew and Moustaki(2011)]{bartholomew2011}
D.~Bartholomew and I.~Moustaki.
\newblock \emph{Latent Variable models and Factor Analysis: A Unified
  Approach}.
\newblock Wiley, third edition, 2011.

\bibitem[Carvalho et~al.(2010)Carvalho, Polson, and Scott]{horseshoe10}
C.~Carvalho, N.~Polson, and J.~Scott.
\newblock The horseshoe estimator for sparse signals.
\newblock \emph{Biometrika}, 97:\penalty0 465--480, 2010.

\bibitem[Carvalho et~al.(2009)Carvalho, Polson, and
  Scott]{carvalho2009handling}
C.~M. Carvalho, N.~G. Polson, and J.~G. Scott.
\newblock Handling sparsity via the horseshoe.
\newblock In \emph{International Conference on Artificial Intelligence and
  Statistics}, pages 73--80, 2009.

\bibitem[Chamberlain and Imbens(1996)]{chamberlain1996hierarchical}
G.~Chamberlain and G.~Imbens.
\newblock Hierarchical {B}ayes models with many instrumental variables, 1996.

\bibitem[Chan and Tobias(2014)]{chan2014imperfect}
J.~Chan and J.~Tobias.
\newblock Priors and posterior computation in linear endogenous variable models
  with imperfect instruments.
\newblock \emph{Journal of Applied Econometrics}, 2014.

\bibitem[Chao and Phillips(1998)]{chao1998posterior}
J.~C. Chao and P.~C. Phillips.
\newblock Posterior distributions in limited information analysis of the
  simultaneous equations model using the {J}effreys prior.
\newblock \emph{Journal of Econometrics}, 87\penalty0 (1):\penalty0 49--86,
  1998.

\bibitem[Clyde et~al.(1996)Clyde, Desimone, and
  Parmigiani]{clyde1996prediction}
M.~Clyde, H.~Desimone, and G.~Parmigiani.
\newblock Prediction via orthogonalized model mixing.
\newblock \emph{Journal of the American Statistical Association}, 91\penalty0
  (435):\penalty0 1197--1208, 1996.

\bibitem[Conley et~al.(2008)Conley, Hansen, McCulloch, and
  Rossi]{conley2008semi}
T.~G. Conley, C.~B. Hansen, R.~E. McCulloch, and P.~E. Rossi.
\newblock A semi-parametric {B}ayesian approach to the instrumental variable
  problem.
\newblock \emph{Journal of Econometrics}, 144\penalty0 (1):\penalty0 276--305,
  2008.

\bibitem[Conley et~al.(2012)Conley, Hansen, and Rossi]{conley2012plausibly}
T.~G. Conley, C.~B. Hansen, and P.~E. Rossi.
\newblock Plausibly exogenous.
\newblock \emph{Review of Economics and Statistics}, 94\penalty0 (1):\penalty0
  260--272, 2012.

\bibitem[Cox(1968)]{Cox}
D.~Cox.
\newblock Notes on some aspects of regression analysis.
\newblock \emph{Journal of the Royal Statistical Society Series A},
  131:\penalty0 265--279, 1968.

\bibitem[Dreze(1976)]{dreze1976bayesian}
J.~H. Dreze.
\newblock Bayesian limited information analysis of the simultaneous equations
  model.
\newblock \emph{Econometrica: Journal of the Econometric Society}, pages
  1045--1075, 1976.

\bibitem[Fazel(2002)]{fazel2002matrix}
M.~Fazel.
\newblock \emph{Matrix rank minimization with applications}.
\newblock PhD thesis, Stanford University, 2002.

\bibitem[Frisch(1934)]{frisch}
R.~Frisch.
\newblock Statistical confluence analysis by means of complete regression
  systems.
\newblock Technical Report~5, University of Oslo, Economic Institute, 1934.

\bibitem[Geweke(1996)]{geweke1996bayesian}
J.~Geweke.
\newblock Bayesian reduced rank regression in econometrics.
\newblock \emph{Journal of Econometrics}, 75\penalty0 (1):\penalty0 121--146,
  1996.

\bibitem[Grant and Boyd(2008)]{gb08}
M.~Grant and S.~Boyd.
\newblock Graph implementations for nonsmooth convex programs.
\newblock In V.~Blondel, S.~Boyd, and H.~Kimura, editors, \emph{Recent Advances
  in Learning and Control}, Lecture Notes in Control and Information Sciences,
  pages 95--110. Springer-Verlag Limited, 2008.

\bibitem[Grant and Boyd(2013)]{cvx}
M.~Grant and S.~Boyd.
\newblock {CVX}: Matlab software for disciplined convex programming, version
  2.0 beta.
\newblock \url{http://cvxr.com/cvx}, Sept. 2013.

\bibitem[Griffin and Brown(2012)]{Griffin12}
J.~Griffin and P.~Brown.
\newblock Structuring shrinkage: some correlated priors for regression.
\newblock \emph{Biometrika}, 99:\penalty0 481--487, 2012.

\bibitem[Groen and Kapetanios(2009)]{groen2009parsimonious}
J.~J. Groen and G.~Kapetanios.
\newblock Parsimonious estimation with many instruments.
\newblock \emph{Federal Reserve Bank of New York, Staff Report}, \penalty0
  (386), 2009.

\bibitem[Hahn and Hansen(2011)]{hahn2011parameter}
J.~Hahn and K.~Hansen.
\newblock Parameter orthogonalization and {B}ayesian inference with many
  instruments.
\newblock \emph{Economics Letters}, 112\penalty0 (2):\penalty0 207--209, 2011.

\bibitem[Hahn et~al.(2013)Hahn, Carvalho, and Mukherjee]{hahn2013partial}
P.~Hahn, C.~M. Carvalho, and S.~Mukherjee.
\newblock Partial factor modeling: predictor-dependent shrinkage for linear
  regression.
\newblock \emph{Journal of the American Statistical Association}, 108\penalty0
  (503):\penalty0 999--1008, 2013.

\bibitem[Jolliffe(1982)]{jolliffe82}
I.~T. Jolliffe.
\newblock A note on the use of principal components in regression.
\newblock \emph{Journal of the Royal Statistical Society, Series C},
  31\penalty0 (3):\penalty0 300--303, 1982.

\bibitem[Kapetanios and Marcellino(2010)]{kapetanios2010factor}
G.~Kapetanios and M.~Marcellino.
\newblock Factor-{GMM} estimation with large sets of possibly weak instruments.
\newblock \emph{Computational Statistics \& Data Analysis}, 54\penalty0
  (11):\penalty0 2655--2675, 2010.

\bibitem[Kleibergen and Zivot(2003)]{kleibergen2003bayesian}
F.~Kleibergen and E.~Zivot.
\newblock Bayesian and classical approaches to instrumental variable
  regression.
\newblock \emph{Journal of Econometrics}, 114\penalty0 (1):\penalty0 29--72,
  2003.

\bibitem[Liang et~al.(2007)Liang, Mukherjee, and West]{Liang2007}
F.~Liang, S.~Mukherjee, and M.~West.
\newblock The use of unlabeled data in predictive modeling.
\newblock \emph{Statistical Science}, 22\penalty0 (2):\penalty0 189--205, 2007.

\bibitem[Liang et~al.(2008)Liang, Paulo, Molina, Clyde, and
  Berger]{Liangetal08}
F.~Liang, R.~Paulo, G.~Molina, M.~Clyde, and J.~Berger.
\newblock Mixtures of g priors for {B}ayesian variable selection.
\newblock \emph{Journal of the American Statistical Association}, 103:\penalty0
  410--423, 2008.

\bibitem[Lindley and El-Sayed(1968)]{LindleyElSayyad1968}
D.~Lindley and G.~El-Sayed.
\newblock The {B}ayesian estimation of a linear functional relationship.
\newblock \emph{Journal of the Royal Statistical Society. Series B},
  30:\penalty0 190--202, 1968.

\bibitem[Lopes and Polson(2014)]{lopes2014bayesian}
H.~F. Lopes and N.~G. Polson.
\newblock Bayesian instrumental variables: priors and likelihoods.
\newblock \emph{Econometric Reviews}, 33\penalty0 (1-4):\penalty0 100--121,
  2014.

\bibitem[Lopes et~al.(2008)Lopes, Salazar, and Gamerman]{lopes2008spatial}
H.~F. Lopes, E.~Salazar, and D.~Gamerman.
\newblock Spatial dynamic factor analysis.
\newblock \emph{Bayesian Analysis}, 3\penalty0 (4):\penalty0 759--792, 2008.

\bibitem[Ludvigson and Ng(2007)]{ludvigson2007empirical}
S.~C. Ludvigson and S.~Ng.
\newblock The empirical risk--return relation: A factor analysis approach.
\newblock \emph{Journal of Financial Economics}, 83\penalty0 (1):\penalty0
  171--222, 2007.

\bibitem[Maruyama and George(2011)]{maruyamaGeorge2011}
Y.~Maruyama and E.~I. George.
\newblock Fully {B}ayes factors with a generalized {g}-prior.
\newblock \emph{The Annals of Statistics}, 39\penalty0 (5):\penalty0
  2740--2765, 2011.

\bibitem[Murray et~al.(2013)Murray, Dunson, Carin, and
  Lucas]{murray2013bayesian}
J.~S. Murray, D.~B. Dunson, L.~Carin, and J.~E. Lucas.
\newblock Bayesian {G}aussian copula factor models for mixed data.
\newblock \emph{Journal of the American Statistical Association}, 108\penalty0
  (502):\penalty0 656--665, 2013.

\bibitem[Ng and Bai(2009)]{ng2009selecting}
S.~Ng and J.~Bai.
\newblock Selecting instrumental variables in a data rich environment.
\newblock \emph{Journal of Time Series Econometrics}, 1\penalty0 (1), 2009.

\bibitem[Ning et~al.(2013)Ning, Georgiou, Tannenbaum, and Boyd]{ning2013linear}
L.~Ning, T.~T. Georgiou, A.~Tannenbaum, and S.~P. Boyd.
\newblock Linear models based on noisy data and the {F}risch scheme.
\newblock \emph{arXiv preprint arXiv:1304.3877}, 2013.

\bibitem[Polson and Scott(2012)]{PolsonScott2012}
N.~Polson and J.~Scott.
\newblock Local shrinkage rules, {L}evy processes and regularized regression.
\newblock \emph{Journal of the Royal Statistical Society, B}, 74(2):\penalty0
  287--311, 2012.

\bibitem[Rossi et~al.(2006)Rossi, Allenby, and McCulloch]{BayesMarketing}
P.~E. Rossi, G.~M. Allenby, and R.~McCulloch.
\newblock \emph{{B}ayesian statistics and marketing}.
\newblock Series in Probability and Statistics. Wiley, 2006.

\bibitem[Shapiro(1982)]{Shapiro1982weighted}
A.~Shapiro.
\newblock Weighted minimum trace factor analysis.
\newblock \emph{Psychometrika}, 47\penalty0 (3):\penalty0 243--264, 1982.

\bibitem[Spearman(1904)]{spearman1904}
C.~Spearman.
\newblock General intelligence, objectively determined and measured.
\newblock \emph{American Journal of Psychology}, 15:\penalty0 201--293, 1904.

\bibitem[West(2003)]{west2003}
M.~West.
\newblock {B}ayesian factor regression models in the ``large p, small n''
  paradigm.
\newblock In J.~M. Bernardo, M.~Bayarri, J.~Berger, A.~Dawid, D.~Heckerman,
  A.~Smith, and M.~West, editors, \emph{Bayesian Statistics 7}, pages 723--732.
  Oxford University Press, 2003.

\bibitem[Yogo(2004)]{yogo2004estimating}
M.~Yogo.
\newblock Estimating the elasticity of intertemporal substitution when
  instruments are weak.
\newblock \emph{Review of Economics and Statistics}, 86\penalty0 (3):\penalty0
  797--810, 2004.

\bibitem[Zellner(1986)]{zellner}
A.~Zellner.
\newblock On assessing prior distributions and {B}ayesian regression analysis
  with g-prior distributions.
\newblock In \emph{Bayesian Inference and Decision Techniques: Essays in Honor
  of Bruno de Finetti}, pages 233--243. Amsterdam: North-Holland, 1986.

\end{thebibliography}
\newpage
\begin{appendix}

\section{Data description}
The table below lists the macroeconomic indicators used in our instruments matrix by mnemonic and accompanied by brief abbreviated descriptions.  These data came originally from the now-defunct DRI-Global Insight, Basic Economics Database which has been subsumed by the IHS Economics \& Country Risk database.  Compare to the table in Appendix A.1 in \cite{ludvigson2007empirical}, of which our list is a subset (some of the series are no longer kept).  Following those authors, we apply the following data transformations (DT), indicated by: 1=no transformation; 2 = first difference; 3 = log first difference. \\
\tiny
\begin{longtable}{lll}
Category/Name & DT & Description\\
FX &  & \\ \hline
BPAUS & 2 & U.S. ASSETS ABROAD (NET) \\
BPB & 2 & BALANCE ON MERCHANDISE TRADE    \\
GDFXFC & 3 & CHAIN-TYPE QUANTITY INDEX - EXPORTS OF GOODS AND SERV  \\
GNET & 2 & NET EXPORTS OF GOODS AND SERV     \\
GRFIW & 3 & RECEIPT FACTOR INCOME FROM REST OF WORLD    \\
GXIM & 1 & \% CHG FRM PRECEDING PERIOD: IMPORTS      \\
GXMDQF & 3 & EXPORTS-DURABLE GOODS \\
GXMNQF & 3 & EXPORTS-NONDURABLE GOODS       \\
GXMQF & 3 & EXPORTS-GOODS     \\
GDFMFC & 3 & CHAIN-TYPE QUANTITY INDEX - IMPORTS OF GOODS AND SERV   \\
\\
Consumption  &  &              \\ \hline
GDFCDC & 3 & CHAIN-TYPE QUANTITY INDEX - PCE, DURABLE GOODS     \\
GXDAQF & 3 & AUTO OUTPUT-EXPORTS     \\
GXPC & 1 & \% CHG FROM PRECEDING PERIOD:PERSONAL CONSUMPTION EXPENDS  \\
GDFCFC & 3 & CHAIN-TYPE QUANTITY INDEX - PERSONAL CONSUMPTION EXPENDITURES     \\
\\
Prices &  &              \\ \hline
GD & 3 & IMPLICIT PR DEFLATOR: GROSS NATIONAL PRODUCT      \\
GDC & 3 & IMPLICIT PR DEFLATOR: PERSONAL CONSUMPTION EXPENDITURES      \\
GDCD & 3 & IMPLICIT PR DEFLATOR: DURABLE GOODS,PCE       \\
GDCN & 3 & IMPLICIT PR DEFLATOR: NONDURABLE GOODS,PCE       \\
GDCS & 3 & IMPLICIT PR DEFLATOR: SERVICES, PCE       \\
GDEX & 3 & IMPLICIT PR DEFLATOR: EXPORTS OF GDS \& SERV   \\
GDEXIM & 3 & TERMS OF TRADE         \\
GDFCC & 3 & CHAIN-TYPE PRICE INDEX - PERSONAL CONSUMPTION EXPENDITURES     \\
GDFCNC & 3 & CHAIN-TYPE PRICE INDEX - PCE, NONDURABLE GOODS     \\
GDFCSC & 3 & CHAIN-TYPE PRICE INDEX - PCE, SERVICES      \\
GDFDCF & 3 & CHAIN-TYPE PRICE INDEX - NATL DEFENSE EXPENDITURES \& GROSS INVESTMENT \\
GDFDFC & 3 & CHAIN-TYPE PRICE INDEX - PCE, DURABLE GOODS     \\
GDFDPC & 3 & CHAIN-TYPE PRICE INDEX- PRODUCERS' DURABLE EQUIPMENT      \\
GDFEXC & 3 & CHAIN-TYPE PRICE INDEX - EXPORTS OF GOODS AND SERVICES   \\
GDFGEC & 3 & CHAIN-TYPE PRICE INDEX - GOVT CONSUMPTION EXPENDITURES \& GROSS INV \\
GDFGFC & 3 & CHAIN-TYPE PRICE INDEX - FED CONSUMPTION EXPEND \& GROSS INVESTMENT \\
GDFGOC & 3 & CHAIN-TYPE PRICE INDEX - NONDEF CONS EXPENDITURES \& GROSS INVESTMENT \\
GDFGSC & 3 & CHAIN-TYPE PRICE INDEX - S\&L CONSUMPTION EXPEND \& GROSS INVESTMENT \\
GDFICF & 3 & CHAIN-TYPE PRICE INDEX - PRIVATE FIXED INVESTMENT     \\
GDFIMC & 3 & CHAIN-TYPE PRICE INDEX - IMPORTS OF GOODS AND SERV   \\
GDFIRC & 3 & CHAIN-TYPE PRICE INDEX - RESIDENTIAL       \\
GDFISC & 3 & CHAIN-TYPE PRICE INDEX - NONRESIDENTIAL STRUCTURES      \\
GDFNRC & 3 & CHAIN-TYPE PRICE INDEX - NONRESIDENTIAL       \\
GDGF & 3 & IMPLICIT PR DEFLATOR: FED GOV'T PURCH OF GDS \& SERV  \\
GDIS & 3 & IMPLICIT PR DEFLATOR: PRIVATE NONRESINDENTIAL STRUCTURES      \\
LBGDPU & 3 & IMPLICIT PRICE DEFLATOR: NONFARM BUSINESS     \\
\\
Fixed Investment &  &              \\ \hline
GFINO & 3 & FIXED INVEST:PRODUCER DURABLE EQUIP      \\
GXIFN & 1 & \% CHG FRM PRECEDING PERIOD:NONRESIDENTIAL FIXED INVESTMENT   \\
GXIFR & 1 & \% CHG FRM PRECEDING PERIOD:RESIDENTIAL FIXED INVESTMENT     \\
GXIPD & 1 & \% CHG FRM PRECEDING PERIOD: NONRESID PRODUCERS' DUR EQUIP    \\
GXIS & 1 & \% CHG FRM PRECEDING PERIOD: NONRESIDENTIAL STRUCTURES    \\
GXPI & 1 & \% CHG FRM PRECEDING PERIOD:GROSS PRIV DOM INVESTMENT    \\
GDFFIC & 3 & CHAIN-TYPE QUANTITY INDEX - PRIVATE FIXED INVESTMENT     \\
GDFIFC & 3 & CHAIN-TYPE QUANTITY INDEX - GROSS PRIVATE DOMESTIC INVESTMENT    \\
\\
Output \& Income &  &              \\ \hline
GDFDEC & 3 & CHAIN-TYPE QUANTITY INDEX - NATL DEF EXPENDITURES \& GROSS INVESTMENTS \\
GDFEOC & 3 & CHAIN-TYPE QUANTITY INDEX - NONDEF CONS EXPEND \& GROSS INVESTMENT \\
GDFFGC & 3 & CHAIN-TYPE QUANTITY INDEX - FED CONSUMPTION EXPEND \& GROSS INVESTMENT \\
GDFGGC & 3 & CHAIN-TYPE QUANTITY INDEX - GOVT CONSUMPTION EXPENDITURES \& GROSS  \\
GDFGLC & 3 & CHAIN-TYPE QUANTITY INDEX - S\&L CONSUMPTION EXPEND \& GROSS INVESTMENT \\
GDFINC & 3 & CHAIN-TYPE QUANTITY INDEX - NONRESIDENTIAL       \\
GDFNFC & 3 & CHAIN-TYPE QUANTITY INDEX - PCE, NONDURABLE GOODS     \\
GDFPDC & 3 & CHAIN-TYPE QUANTITY INDEX - PRODUCERS' DURABLE EQUIPMENT     \\
GDFRFC & 3 & CHAIN-TYPE QUANTITY INDEX - RESIDENTIAL       \\
GDFSFC & 3 & CHAIN-TYPE QUANTITY INDEX - PCE, SERVICES      \\
GDFSTC & 3 & CHAIN-TYPE QUANTITY INDEX - NONRESIDENTIAL STRUCTURES      \\
GPY & 3 & PERSONAL INCOME, TOTAL         \\
GWY & 3 & NAT'L INCOME: WAGES AND SALARIES       \\
GXNP & 1 & \% CHANGE FROM PRECEDING PERIOD, GNP    \\
GXSAV & 3 & PERSN'L INCOME: PERS SAVING RATE, GPSAV AS \% OF GYD  \\
GXYD & 1 & \% CHG FRM PRECEDING PERIOD: DISP. PERSONAL INCOME  \\
GYDPCQ & 3 & DISPOSABLE PERSONAL INCOME PER CAPITA IN CHAINED   \\
GYFIR & 3 & GY BY IND DIV: FINANCE, INSUR AND REAL ESTATE   \\
GYGGE & 3 & GY BY IND DIV: GOV'T AND GOV'T ENTERPRISES    \\
GYM & 3 & GY BY IND DIV: MANUFACTURING INDUSTRY      \\
GYMD & 3 & GY BY IND DIV: DURABLE GOODS MANUFACTURING INDUSTRY    \\
GYMN & 3 & GY BY IND DIV: NONDURABLE GOODS MANUFACTURING INDUSTRY    \\
GYS & 3 & GY BY IND DIV: SERVICE INDUSTRIES      \\
GYT & 3 & GY BY IND DIV: TRANSPORTATION INDUSTRY      \\
GYUT & 3 & GY BY IND DIV: ELECTRIC, GAS AND SANITARY SEW INDUSTRY  \\
\\
Sales, Orders, Purchases &  &              \\ \hline
GXNPD & 1 & GROSS DOM PURCH      \\
GXNS & 1 & FINAL SALES OF DOM PROD     \\
GXNSD & 1 & FINAL SALE TO DOM PURCH     \\
LBOUT & 3 & OUTPUT PER HOUR ALL PERSONS  \\
LBOUTU & 3 & OUTPUT PER HOUR ALL PERSONS: NONFARM BUSINESS    \\
\end{longtable}

\end{appendix}

\end{document}